\documentclass[3p]{elsarticle}

\usepackage{hyperref}

\usepackage{amssymb}
\usepackage{bm}
\usepackage{dcolumn}
\usepackage{amssymb}
\usepackage{amsmath}
\usepackage{graphicx}
\usepackage{booktabs}
\usepackage{multirow}
\usepackage{setspace}
\usepackage{array}
\usepackage{threeparttable}
\usepackage{txfonts}

\journal{APP}

\begin{document}

\begin{frontmatter}



\newcommand*{\PKU}{School of Physics and State Key Laboratory of Nuclear Physics and
Technology, Peking University, Beijing 100871,
China}
\newcommand*{\CIC}{Collaborative Innovation Center of Quantum Matter, Beijing, China}
\newcommand*{\CHEP}{Center for High Energy Physics, Peking University, Beijing 100871, China}
\newcommand*{\CHPS}{Center for History and Philosophy of Science, Peking University, Beijing 100871,
China}

\title{Lorentz violation from gamma-ray bursts}

\author[a]{Shu Zhang}
\author[a,b,c,d]{Bo-Qiang Ma\corref{cor1}}

\address[a]{\PKU}
\address[b]{\CIC}
\address[c]{\CHEP}
\address[d]{\CHPS}
\cortext[cor1]{Corresponding author \ead{mabq@pku.edu.cn}}

\begin{abstract}
The constancy of light speed is a basic assumption in Einstein's special relativity, and consequently the Lorentz invariance is a
fundamental symmetry of space-time in modern physics. However, it is speculated that the speed of light becomes energy-dependent due to the Lorentz invariance
violation~(LV) in various new physics theories.
We analyse the data of the energetic photons from the gamma-ray bursts (GRBs) by the Fermi Gamma-Ray Space Telescope,
and find more events to support the energy dependence in the light speed with both linear and quadratic form corrections.
We provide two scenarios to understand all the new-released Pass~8 data of bright GRBs by the Fermi-LAT Collaboration, with predictions from such scenarios being testable by future detected GRBs.
\end{abstract}


\end{frontmatter}

Since the launch of the Fermi Gamma-Ray Space Telescope (Fermi) in 2008, the Fermi Large Area Telescope~(LAT) on board the Fermi
has detected a number of gamma-ray bursts~(GRBs) with over 10~GeV photons~\cite{catalog}.
These GRBs with measured redshifts and detected energetic photons have been deemed as useful for probing the Lorentz violation~(LV) physics~\citep{GA1998},
as Lorentz invariance is predicted to be broken at the Planck scale~($E \sim E_{\rm Pl}=\sqrt{\hbar c^5/G} \simeq 1.22\times 10^{19} ~\mathrm{GeV}$)~\citep{AmelinoCamelia:2008qg,LVreview}.
As several brightest GRBs with different redshifts have been detected by LAT for now~\citep{catalog,130427ALAT},
it is necessary to check the possibility of finding evidence for or against Lorentz violation effect in high energy photons from these GRBs.

As a consequence of Lorentz violation, the speed of light could have an energy dependence from the expression $v=\partial E/\partial p$.
In a general model-independent form, the modified dispersion relation of photon can be expressed by the leading term of the Taylor series as
\begin{equation}\label{eq:spectral}
  E^2=p^2c^2\left[1-s_\mathrm{n}\left( \frac{pc}{E_{\rm LV,n}}\right)^n\right],
\end{equation}
which corresponds to a modified light speed
\begin{equation}
 v(E)=c\left[1-s_\mathrm{n}\frac{n+1}{2} \left( \frac{E}{E_{\rm LV,n}}\right)^n \right],
\end{equation}
where n=1~or n=2 corresponds to linear or quadratic energy dependence, and $s_n=\pm1$ is the sign of the LV correction.
If $s_n=+1~(s_n=-1)$, the high energy photons travel in vacuum slower~(faster) than the low energy photons. The $n$th-order Lorentz violation scale is characterized by the Lorentz violation parameter $E_{\rm LV, n}$.
Because of the spectral dispersion, two GRB photons emitted simultaneously by the source would
arrive on Earth with a time delay~($\Delta t$) if they have different energies .
With the magnification of the cosmological distances of the GRBs and the high energies of these photons, the time delay ($\Delta t$) caused by the effect of Lorentz violation would be measurable~\citep{GA1998}.
Taking account of the cosmological expansion, we write the formula of the time delay as~\citep{Jacob:2008bw}:
\begin{equation}\label{eq:Deltat}
  \Delta t=t_\mathrm{h}-t_\mathrm{l}=s_\mathrm{n}\frac{1+n}{2H_0}
  \frac{E_\mathrm{h}^n-E_\mathrm{l}^n}{E_{\rm LV,n}^n}
  \int^z_0
  \frac{(1+z')^n dz'}{\sqrt{\Omega_m (1+z')^3+\Omega_\Lambda}}.
\end{equation}
Here, $t_\mathrm{h}$ is the arrival time of the high energy photon, and $t_\mathrm{l}$ is the arrival time of the low energy photon,
with $E_\mathrm{h}$ and $E_\mathrm{l}$ being the photon energies measured by the LAT.
We adopt the cosmological constants $[\Omega_m, \Omega_\Lambda]=[0.272,0.728]$ determined by the latest data from WMAP~\citep{WMAP} and the Hubble constant $H_0=73.8\pm2.4 ~\mathrm{km\ s^{-1}\ Mpc^{-1}}$ measured by the Hubble Space Telescope recently~\citep{hubble}.

However, there are big ambiguities in applying Eq.~(\ref{eq:Deltat}) to analyse the data, because there are also time delays
due to the intrinsic properties of GRBs.
Some works constrained the quantum gravity~(QG) scale $E_{\rm QG}>E_{\rm Pl}$ with some statistic methods~\citep{discan,Vasileiou} and the sharp shape showed in the light curve of the short GRB~090510~\citep{nature090510,PRL090510}, but such big constraints have no support from the other long GRBs.
On the other hand, there are also works~\citep{shao,Amelino} supporting
the Lorentz violation at the energy scale of $10^{17}$~GeV by including the intrinsic time lags of the high energy photons emitted at the GRB source.
In our work, we will analyse all the data of bright GRBs newly released by the Fermi-LAT Collaboration~\citep{pass8}, to consider the Lorentz violation effect in the time delays by including also the intrinsic time lag, following the approach in Refs.~\citep{shao,e06}.

In Eq.~(\ref{eq:Deltat}), a larger energy difference could cause a larger detectable time delay,
so we first analyse the data of the photons with highest energies from 6 bright GRBs,
which are GRBs 080916C, 090510~(short GRB), 090902B, 090926A, 100414A, and 130427A.
Among these GRBs, only GRB 090510 is a short GRB (with duration time~$<$~2~s), while the rest are long GRBs~(with duration time~$>$~2~s).
The photon data of our candidates are collected from the latest Pass~8 Fermi-LAT event reconstruction~\citep{pass8}.
As the data of GRB 130427A have not been re-analyzed with Pass~8, we use the Pass~7 data downloaded from the Fermi-LAT Data Server.
In addition, the 4 highest energy photons associated with GRB 130427A  have comparable energies,
but the first 73-GeV photon which arrived 18~s after the onset of this GRB will be analyzed together with the photons of the earlier bright GRBs.
As the redshift of GRB 130427A ($z=0.34$)~\citep{130427Az} is smaller than those of other considered GRBs,
the large time delays~($>200$ s) measured in the rest 3 high energy gamma-rays are less likely to be the consequence of the Lorentz violation effect,
but more possible to be caused by the large intrinsic time lags between these high energy gamma-rays and the trigger gamma-rays.
For GRB 080916C, both the 2 high energy photons in Pass~8 data are included in our analysis due to the large distance~($z=4.35$)~\citep{080916Cz},
as both of these events could show detectable LV-induced time lags.

The values of $E_{\rm LV,1}$ listed in Table~\ref{tab:grbs} are calculated directly by using Eq.~(\ref{eq:Deltat}),
where $t_{\rm l}$ is the trigger time of the GRB detected by the Gamma-Ray Burst Monitor~(GBM)~\citep{GBM},
and $t_{\rm h}$ is the arrival time of the high energy photon.
Because the photons arrived at the trigger time have low energies at the order of 100~keV~\citep{catalog},
we consider $E_{\rm l}$ in Eq.~(\ref{eq:Deltat}) as 0 approximately in our calculation.
From Table~\ref{tab:grbs} we see that the values of $E_{\rm LV,1}$ are all around the order of $10^{17}$~GeV except for the short GRB~090510, which
leads to a large $E_{\rm LV,1}\sim10^{19}$~GeV~\citep{Xiao:2009xe}.
We notice the slight difference in $E_{\rm LV,1}$'s for long GRBs and the large gap between these $E_{\rm LV,1}$'s and that for short GRB~090510. In fact, $E_{\rm LV,1}$'s
in Table~\ref{tab:grbs} are calculated under the assumption that both the high energy photons and the onset low energy photons are emitted at the source
simultaneously. We may attribute the difference to an intrinsic time lag of the high energy photon emitted at the source as compared with the emission time
of the low energy photons. Therefore the slight difference for long GRBs might imply that these high energy photons have comparable intrinsic emission time,
while the high energy photon of short GRB~090510 has a quite different intrinsic emission time.
Taking into account the intrinsic emission time $t_{\rm in}$  of high energy photons, we write the observed time delay between the onset of GRB (trigger time) photons and the high energy photons as
\begin{equation}\label{eq:tobs}
 t_{\rm obs}=  t_{\rm LV}+(1+z) t_{\rm in},
\end{equation}
where $t_{\rm LV}$ is now the time lag caused by the Lorentz violation, i.e., $\Delta t$ in Eq.~(\ref{eq:Deltat}).

\begin{threeparttable}

  \caption{The data of the GRBs with high energy photons and known redshifts.}
    \begin{tabular}{cccccp{23mm}<{\centering}cp{26mm}<{\centering}}
    \hline
    \hline
   GRB         & z           & $t_{\rm obs}$~(s) & $E_{\rm obs}$~(GeV) & $E_{\rm in}$~(GeV) & $E_{\rm LV,1}$ ($\times~10^{17}$~GeV) & $\frac{t_{\rm obs}}{1+z}$~(s) & $K_{\rm 1}$ ($\times 10^{18}~\mathrm{s}~\cdot$~GeV) \\
    \hline
    080916C(1)  & $4.35\pm0.15$ & 16.545      & 12.4        & 66.3        & $13.9\pm 1.7$        & 3.092       & 4.30  \\
    090926A     & $ 2.1071\pm0.0001$ & 24.835      & 19.5        & 60.6        & $7.8\pm 0.8$         & 7.993       & 6.23  \\
    100414A     & $1.368$ & 33.365      & 29.7        & 70.3        & $5.8\pm 0.6$         & 14.090      & 8.22  \\
    130427A$^a$     & $0.3399 \pm 0.0002$ & 18.644      & 72.6        & 97.3        & $6.0\pm 0.7$         & 13.915      & 8.32  \\
    090902B     & 1.822 & 81.746      & 39.9        & 112.6       & $4.2\pm 0.5$         & 28.967      & 12.24  \\
    \hline
    090510      & $0.903 \pm 0.003$   & 0.828       & 29.9        & 56.9        & $155\pm 17$       & 0.435       & 6.75  \\
    080916C(2)  & $4.35\pm0.15$ & 40.509      & 27.4        & 146.6       & $12.6 \pm 1.4$        & 7.572       & 9.51  \\
    \hline
     & & 11.671      & 11.9        & 33.6        & $8.8 \pm 1.0$         & 4.136       & 3.65  \\
            &    & 14.166      & 14.2        & 40.1        & $8.7 \pm 1.0$         & 5.020       & 4.36  \\
        090902Bs    &  1.822  & 26.168      & 18.1        & 51.1        & $6.0 \pm 0.7$         & 9.273       & 5.55  \\
            &    & 42.374      & 12.7        & 35.8        & $2.6 \pm 0.3$         & 15.016      & 3.90  \\
            &    & 45.608      & 15.4        & 43.5        & $2.9 \pm 0.3$        & 16.162      & 4.72  \\
    \hline
    \hline
    \end{tabular}%

    \begin{tablenotes}
      \item  $^a$The data of this GRB are from the Pass~7 LAT reconstruction.
        The references for the redshifts of the GRBs are \cite{080916Cz}(GRB~080916C), \cite{090510z}(GRB~090510), \cite{090902Bz}(GRB~090902B), \cite{090926Az}(GRB~090926A), \cite{100414Az}(GRB~100414A), and \cite{130427Az}(GRB~130427A).
        $t_{\rm obs}$ is the arrival time after the onset of the GRBs, $E_{\rm obs}$ is the measured energy of the photon,
  $E_{\rm in}$ is the intrinsic energy at the source of the GRBs,
  and $E_{\rm LV,1}$ is the Lorentz violation parameter of the linear LV model without considering the intrinsic time lag.
  The standard errors of $E_{\rm LV,1}$'s are calculated with the consideration of the energy resolution of LAT~\cite{LAT2012} and the uncertainties of the cosmological parameters and the redshifts.
  $K_{\rm 1}$ is the Lorentz violation factor with a unit as~(s~$\cdot$~GeV)
    \end{tablenotes}
  \label{tab:grbs}%
\end{threeparttable}%

\begin{figure}
   \centering
  \includegraphics[width=90mm]{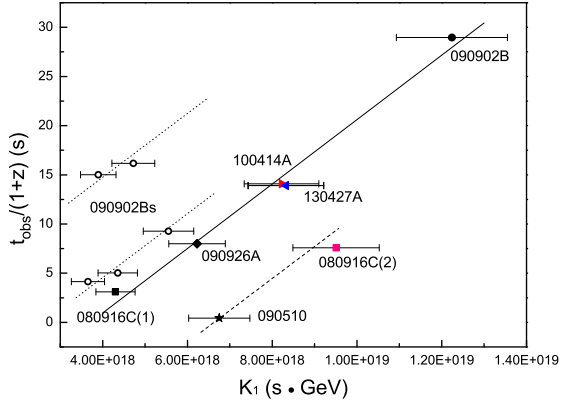}
\caption{The $\frac{\Delta t_{\rm obs}}{1+z}$ versus $K_{\rm 1}$ plot for high energy photons from GRBs with Eq.~(\ref{eq:fit}).
The slopes of the lines are $1/E_{\rm LV,1}$ with $s_{1}=+1$. For the 5 gamma-ray events in the main straight line (black solid line), the intercept is $t_{\rm in}=-12.1\pm1.7$~s. The lower line (dash line) is a straight line passing through the point of the short GRB~090510 (marked by $\star$).
The slope of this line is the same as that of the main line, and the intercept is $t_{\rm in}=-21.7$~s.
The dark solid points are the events available in Ref.~\citep{shao}.
This figure includes all the Pass~8 re-analyzed photon events. The rest 5 events of GRB~090902B are added in this figure as the open circles. Two parallel lines (dot lines) passing through the circles show the possibly same $t_{\rm in}$ of these photons on a same line. The intercepts of the two lines are -8.7~s and 1.5~s respectively. The errors of all the points are under the full consideration of the uncertainties about the energy, cosmological parameters and the redshifts. }
\label{fig:all}
\end{figure}

\vspace*{1cm}
Then we re-express Eq.~(\ref{eq:tobs}) as a linear function in a form:
\begin{equation}\label{eq:fit}
  \frac{ t_{\rm obs}}{1+z}=\frac{K_{\rm n}}{E_{\rm LV,n}^n}+t_{\rm in},
\end{equation}
where the Lorentz violation factor reads
\begin{equation}\label{eq:K}
  K_{\rm n}=\frac{1+n}{2H_0}
  \frac{E^n}{1+z}
  \int^z_0
  \frac{(1+z')^n dz'}{\sqrt{\Omega_m (1+z')^3+\Omega_\Lambda}},
\end{equation}
which has a unit as $[~\mathrm{s} \cdot \mathrm{GeV}]$.
To check the possible linear-dependence of the Lorentz violation effect in the data, we draw the $\frac{\Delta t_{\rm obs}}{1+z}$ versus $K_{\rm 1}$ plot for all the high energy photons in Table~\ref{tab:grbs}
in Fig.~\ref{fig:all}, where the $X$ axis is $K_{\rm 1}$ and the $Y$ axis is $\frac{\Delta t_{\rm obs}}{1+z}$.
As we can see from Fig.~\ref{fig:all}, 5 points of photons from 5 long GRBs can be fitted on a straight line, which can be considered as the main line.
The intercept of this line with the $Y$ axis corresponds to the intrinsic time lag $t_{\rm in}$.
This means that the intrinsic time lags $t_{\rm in}$ between the highest energy gamma rays and the GRB onset are approximately the same for these 5 gamma-rays of long GRBs. We consider this line as a support for
a linear Lorentz violation effect at a scale $E_{\rm LV,1}=(3.05\pm0.19) \times 10^{17}$~GeV with the same intrinsic time lag $t_{\rm in}=-12.1\pm1.7$~s.
Including the energy resolution of LAT for over 10 GeV photons as 10\% ~\citep{LAT2012} and the uncertainties of the cosmological parameters and the redshifts to the fitted slope, we get the standard error of
$E_{\rm LV,1}$ as $0.7\times 10^{17}$~GeV. After we consider these errors, the mixing of Pass~7 data and Pass~8 data do not influence the conclusion of the result.
We see that a higher precision is achieved for this result of $E_{\rm LV,1}$ compared to the averaged value $E_{\rm LV,1}=( 7.5\pm3.4 ) \times 10^{17}$~GeV corresponding to the 5 gamma-rays without considering the intrinsic emission time effect in Table~\ref{tab:grbs}.

In the Pass~8 data of GRB~080916C, a 27.4~GeV photon with a 40.5~s time lag has been reconstructed~\citep{pass8}.
The data and the point are shown in Table~\ref{tab:grbs}~(080916C(2)) and Fig.~\ref{fig:all}~(pink square).
We notice that this gamma-ray may have the same $t_{\rm in}$ as that of the gamma-ray in GRB~090510, for the same sloping line passing through the point of GRB~090510 is just near the point of the 27.4-GeV photon.
This point is consistent with that line when a 10\% energy uncertainty of these gamma-rays is considered.
If this is not a coincidence, it might mean that some high energy photons from long GRBs have a similar emission mechanism to that of some short GRBs.
We also notice from Table~\ref{tab:grbs} that the intrinsic energy $E_{\rm in}$ of these events on the main line are comparable to each other, but the 27.4~GeV gamma-ray has an intrinsic energy $E_{\rm in}=147$~GeV, which is much higher than those of the other gamma-rays.
This can be considered as a hint for a distinct emission mechanism
for this photon with the highest intrinsic energy at the GRB source, thus the difference of $t_{\rm in}$ between the 27.4-GeV gamma-ray and those of the other 5 long GRB events in the above main line becomes reasonable.
Meanwhile, GRB~090510 is the first short GRB which has been detected with an initial optical emission~\citep{ApjLswift}.
With the distinct highest intrinsic energy of the gamma-ray from GRB~080916C and this uncommon similarity between GRB~090510 and long GRBs, our conclusion of a possible same emission mechanism in GRB~090510 and GRB~080916C appears reasonable.
On the other hand, the difference of the $t_{\rm in}$'s between the GRBs 0809510, 080916C(2) and the GRBs on the main line does not violate the previous observations on GRBs. We can see from  Ref.~\citep{catalog} that most lightcurves of the gamma-ray bursts have more than one peak, and the peaks of different energy bands are always different. This might imply that the gamma-rays with different energies have different emission time.

Furthermore, 5 gamma-ray events of GRB~090902B with lower energies have been re-analyzed by Pass~8~\citep{pass8}.
We add them in Fig.~\ref{fig:all} as open circles, and notice that they fall on two parallel lines shown in the figure.
This illustrates that the points on a same line are emitted simultaneously from the source. In fact, photons with the same
intrinsic emission time should fall onto a straight line in the $\frac{\Delta t_{\rm obs}}{1+z}$ versus $K_{\rm 1}$ plot,
but photons with different intrinsic emission times should fall on several parallel lines with the same slope as $1/E_{\rm LV,1}$.
In our analysis we require that the other three lines have the same slope as that of the main line
in Fig.~\ref{fig:all}.

We need to distinguish between two kinds of straight lines, i.e., the line with points from multi-GRBs and the line with all points
from a single GRB. We notice that the main line is a multi-GRB line with 5 points from each 5 long GRBs, whereas the two parallel lines
mentioned above are single-GRB lines.
Even though the data selection used to fit the 5 points from 5 GRBs in the main line is significant and reasonable, we still need to check the rationality of our approach by fitting the photon events with other selections.
For example, when we fit all 6 gamma-rays in the same GRB~090902B, we get $E_{\rm LV,1}=(4.2 \pm 1.3) \times 10^{17}$~GeV.
In addition, by combining the 5 events
on the main line with the 2 events on the lower line and fitting
them with one line, the Lorentz violation parameter becomes
$E_{\rm LV,1}=(3.2 \pm 0.9) \times 10^{17}$~GeV.
We can also fit all points in Fig.~\ref{fig:all}, then we get $E_{\rm LV,1}=(5.7\pm 2.5) \times 10^{17}$~GeV.
All the above results are compatible with that of the main line within 1$\sigma$ error range.
Thus we provide a scenario for a comprehensive
understanding of all the 11 gamma-rays in the new Pass~8 release together with an additional gamma-ray from the Pass~7 data of GRB~130427A.
Although such selections of events with a straight line for fitting can produce results with big errors, these big errors
can be served as a support of our approach to analyse the events by different groups with
different intrinsic time lags.
Further more, the several straight lines plotted in Fig.~\ref{fig:all}, especially the multi-GRB lines, can be considered as our predictions.
We also speculate that photons with higher energies may have more chances to be emitted at the GRB source with earlier
intrinsic emission time, as can be supported from the averaged $E_{\rm in}$ for the four lines in Fig.~\ref{fig:all} in an upward order:
$102\pm45$~GeV, $81\pm20$~GeV, $42\pm7$~GeV, $40\pm4$~GeV.
The rationality of our predictions can be tested by more high energy gamma-rays in future detected GRBs.

In fact, the approach in our paper was adopted in Ref.~\citep{shao}. It was shown there, by confronting with high energy gamma-rays from
4 GRBs 080916C, 090510, 090902B and 090926A (dark solid points in Fig.~\ref{fig:all}), that the 3 high energy photons from 3 long GRBs fall on a straight line with $E_{\rm LV,1}=(2.2\pm 0.2) \times 10^{17}$~GeV.
Here we add further 2 gamma-ray events from the new GRBs 100414A (left red triangle) and 130427A (right blue triangle), and replace the data with the latest LAT reconstruction Pass~8.
It is encouraging to see that now we have 5 events from each 5 long GRBs on the main straight line with $E_{\rm LV,1}=(3.05 \pm 0.19) \times 10^{17}$~GeV. The difference between these two results is acceptable if we consider the energy uncertainty of LAT (10\%) for over 10~GeV photons as the source of the system uncertainty. In addition, as the data reconstruction has been updated from Pass~6 in Ref.~\citep{shao} to Pass~8 now, some energies of photons have been corrected. For example, the energy of the highest energy gamma-ray of GRB~090902B has been changed from 33.4~GeV to 39.9~GeV. Fitting the new data of the three GRBs used in Ref.~\citep{shao} we get $E_{\rm LV,1}=(3.03\pm 0.16)\times 10^{17}$~GeV, which is consistent with the current result for 5 events.
Meanwhile, we also find a new event of GRB~080916C to fall on the parallel line passing through the point of the short GRB~090510. Therefore this work can be considered as a natural extension along Ref.~\citep{shao} with more events to support the analysis.

\begin{figure}
     \centering
 \includegraphics[width=90mm]{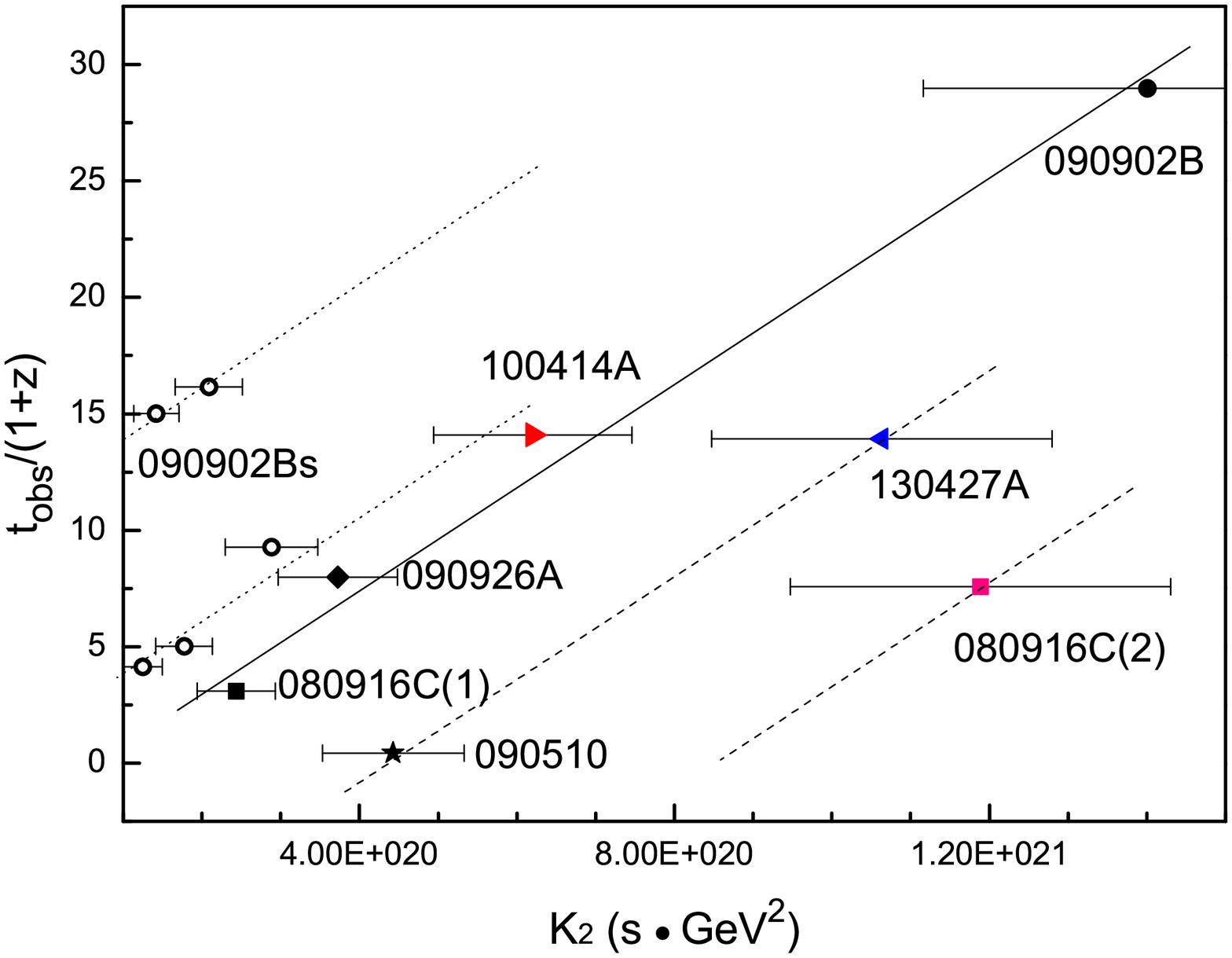}
\caption{The $\frac{\Delta t_{\rm obs}}{1+z}$ versus $K_{\rm 2}$ plot for high energy photons from GRBs with Eq.~(\ref{eq:fit}).
The slopes of the lines are $1/E_{\rm LV,2}^2$ with $s_{1}=+1$ where $E_{\rm LV,2}=(6.8 \pm 1.0)\times 10^{9}$~GeV. For the 3 gamma-ray events (dark solid points) in the main straight line (black solid line), the intercept is $t_{\rm in}=-1.2\pm1.4$~s. The lower line (black dash line) is a straight line passing through the point of the short GRB~090510 (marked by $\star$).
The slope of the line is the same as that of the main line, and the intercept is $t_{\rm in}=-9.0$~s.
The dark solid points are the events available in Ref.~\citep{shao}.
This figure includes all the Pass~8 re-analyzed photon events. The rest 5 events of GRB~090902B are added in this figure as the open circles. Two parallel lines (dot lines) passing through the circles show the possibly same $t_{\rm in}$ of these photons on a same line. The intercepts of the two lines are 1.8~s and 11.9~s respectively. The figure also shows that the highest energy gamma-ray from GRB~080916C has an intrinsic time lag as $t_{\rm in}=-18.3$~s. The errors of all the points are under the full consideration of the uncertainties about the energy, cosmological parameters and the redshifts.}

\label{fig:n2}

\end{figure}

In the case of Lorentz violation correction with a quadratic form, which means $n=2$ in modified spectral dispersion (\ref{eq:spectral}), we draw the $\frac{\Delta t_{\rm obs}}{1+z}$ versus $K_{\rm 2}$ plot for the data of gamma-ray events in Fig.~\ref{fig:n2}.
We notice that the point of the highest energy gamma-ray from GRB~100414A is near the line passing through the 3 high energy gamma-rays from GRB~080916C, GRB~090926A and GRB~090902B. This line was noticed in Ref.~\citep{shao} and we now call this line as the main line for the quadratic case.
The slope of the main line is $1/E_{\rm LV,2}^2$ with $E_{\rm LV,2}=(6.8\pm1.0)\times 10^{9}$~GeV, and
the intercept of this line is the intrinsic time lag with $t_{\rm in}=-1.2\pm1.4$~s. This means that the highest energy gamma-ray from GRB~100414A has
almost the same intrinsic time lag as those of the 3 gamma-rays on the main line.
Then we also notice that the point of the photon event of GRB~130427A now falls on the parallel line passing through the short GRB~090510, indicating that the two gamma-rays have the same intrinsic time lag in this case.
Two parallel lines passing through the 5 points of gamma-rays from GRB~090902B are drawn in Fig.~\ref{fig:n2}, and we also have another line passing through the point of the 27.4-GeV photon from GRB~080916C.  We predict that more points would fall on the lines in Fig.~\ref{fig:n2} if the speed of light is modified by the quadratic form of Lorentz violation correction with $E_{\rm LV,2}=(6.8\pm1.0)\times 10^9$~GeV.
In comparison with the linear case, we provide a different scenario to understand all data of gamma-rays in the new Pass~8 release and the high energy gamma-ray from GRB~130427A in Pass~7.
Because the observed time lag has two components, one is the intrinsic time lag, and the other is the time lag caused by Lorentz violation, the form of the dispersion relation would influence the obtained intrinsic time lags consequently.
At this moment it is still premature to draw a definite conclusion, though the linear case is slightly favored by more events.
The form of Lorentz violation can be checked by comparison between the predictions from the two different scenarios and the future data of high energy gamma-rays from GRBs.

One feature of our scenarios is that some energetic photons might be emitted at the source earlier than the low energy photons
that served as the onset of the GRB by detection. Therefore there could be cases that some high energy photons may have earlier arrival time if the redshifts of the GRBs are not so large.
It is interesting to notice that many GRBs (like GRB~081006, GRB~090217, GRB~090510 and other 9 GRBs) are found to have high energy ($>$0.5~GeV) photons with minus arrival time~\citep{catalog}. Among these GRBs, GRB~090510 and GRB~090227B are short GRBs.
This seems to be in coincidence with our results of negative intrinsic time lag of energetic photons.
Unfortunately, most of these GRBs do not have known redshifts so that we cannot study these cases in detail.
But the existence of high energy photons with minus arrival time could be regarded as a support for a possible mechanism that
the GRBs can emit high energy photons earlier than low energy photons at the source.
Besides the minus arrival time shown in the $>$ 0.5~GeV gamma-rays, the earlier arrival time of higher energy photons are also shown in the 10-1000~keV energy band in long GRBs
for the prompt emissions~\citep{1986}. All these phenomena mentioned above support the existence of the intrinsic time lag we calculated for over 10 GeV gamma-rays.

To further examine our results, we compare them with the limitations set from active galactic nuclei (AGNs) and some individual analysis of GRBs.
The energy-dependent delay of the high energy photons from the AGN Markarian~501 was measured, and it was used to limit the Lorentz violation parameters as $E_{\rm LV,1}>2.1\times 10^{17}$~GeV at the 95\% C.L.~\citep{MKN501}. This result is compatible with our value of $E_{\rm LV,1}$.
In the results of GRBs, the highest energy (31 GeV) photon in GRB~090510 was considered to set the limitation of $E_{\rm LV,1}$ to over $E_{\rm Pl}=1.22\times 10^{19}$~GeV with the assumption of positive intrinsic time lag~\citep{Vasileiou,nature090510}. If we set the intrinsic time lag $t_{\rm in}=0$, we also get this result as showed in Table~\ref{tab:grbs}, where $E_{\rm LV,1}=154.9\times 10^{17}$~GeV for GRB~090510.
Therefore the main reason for the difference between our conclusion and those in Refs.\citep{Vasileiou,nature090510} is that we determine the intrinsic time lag by fitting high energy photons from several GRBs whereas Refs.~\citep{Vasileiou,nature090510} analyzed the high energy photons from an individual GRB~090510 with the assumption of a positive intrinsic time lag.

Before conclusion, we need to discuss possible implications of our results. In the linear case of the Lorentz violation effect, the Lorentz
violation parameter from our result is $E_{\rm LV, 1}=(3.0 \pm 0.7) \times 10^{17}$~GeV. Under a cautious
consideration, we may take this result as the lower limit of the linear Lorentz violation parameter $E_{\rm LV, 1}>2.3 \times
10^{17}$~GeV. Besides Lorentz violation, the matter in the universe can also cause the dispersive effects in the
propagation of light~\citep{darkmatter}. The regularities revealed in the GRB data by our result could be a consequence of the matter effect in space. If we
consider this energy-dependent light speed showed in our analysis as caused by the Lorentz violation effect in vacuum, the
linear model has more chances to express the LV scale as around $E_{\rm LV, 1}$, which is just below the Planck scale $E_{\rm Pl}$. In fact, the Planck scale
is just an estimate for the scale at which the Lorentz invariance might be broken, and the exact value should be determined from experimental evidence rather than a~priori assumption~\cite{Planck-scale}.
On the other hand, $E_{\rm LV, 2}$ in the quadratic model is far from the Planck scale, therefore the quadratic model is only likely to survive
in case that the regularities in the GRB data is due to the cosmological matter effect as suggested in Ref.~\citep{darkmatter}.

In conclusion, we try to understand all energetic gamma-rays of the Pass~8 data by the Fermi-LAT Collaboration
with two scenarios for both linear and quadratic forms of Lorentz violation correction to the light speed.
By adopting an approach with the high energy photons emitting at the GRB source with different intrinsic time compared with the onset time of the low energy gamma-rays, we obtain a linear Lorentz violation parameter $E_{\rm LV,1}=(3.05\pm 0.19) \times 10^{17}$~GeV
by fitting 5 gamma rays from 5 long GRBs.
Such a scale is also supported by fittings of all 6 gamma-rays in GRB~090902B and all 11 gamma-rays in the Pass~8 release plus an additional high energy photon event
in GRB~130427A with the shortest arrival time. As a comparison,
we also obtain a quadratic Lorentz violation parameter $E_{\rm LV,2}=(6.8\pm1.0)\times 10^{9}$~GeV by fitting 3 gamma rays from 3 long GRBs,
with one gamma ray from GRB~100414A falling near the line passing through the above 3 points. The other events of energetic gamma rays
can be considered as with different intrinsic emission time, as shown by several parallel lines in both the two scenarios in Figs.~\ref{fig:all} and \ref{fig:n2}. Therefore we find more events to support the energy dependence in the light speed with both linear and quadratic forms of Lorentz violation correction, and we also suggest predictions that can be clearly tested by more data in future experiments.


{\bf Acknowledgements}
This work is supported by National Natural Science Foundation of China (Grants No.~11021092, No.~10975003, No.~11035003, and No.~11120101004) and the National Fund for Fostering Talents of Basic Science (Grant Nos.~J1103205 and J1103206). It is also supported by the Undergraduate Research Fund of Education Foundation of Peking University.




\begin{thebibliography}{99}
\bibitem{catalog}
M.~Ackermann, et al.[Fermi-LAT Collaboration],
Astrophys.\ J.\ Suppl.\  {209} (2013) 11.




\bibitem{GA1998}
G.~Amelino-Camelia, J.~R.~Ellis, N.~E.~Mavromatos, D.~V.~Nanopoulos, S.~Sarkar,
  Nature  393 (1998) 763.

\bibitem{AmelinoCamelia:2008qg}
For a recent review on quantum gravity theories and the relevant Lorentz violation studies, see, e.g.,
  G.~Amelino-Camelia,
  Living Rev.\ Rel.\  {16} (2013) 5,
    and references therein.



\bibitem{LVreview}
  For a brief review on studing Lorentz violation with very high energy photons, see, e.g.,
  L.~Shao and B.-Q.~Ma,
  Mod.\ Phys.\ Lett.\ A 25 (2010) 3251,
  and references therein.

\bibitem[Zhu et al.(2013)]{130427ALAT}
S.~Zhu { et al.},
GCN Circ. 14471 (2013).





\bibitem{Jacob:2008bw}
  U.~Jacob, T.~Piran,
  JCAP 0801 (2008) 031.


\bibitem{WMAP}
  E.~Komatsu { et al.}[WMAP Collaboration],
   Astrophys.\ J.\ Suppl.\  192 (2011) 18.

\bibitem{hubble}
  A.~G.~Riess et al.,
   Astrophys.\ J.\   730 (2011) 119;
  Erratum-ibid.\   732 (2011) 129.

\bibitem{discan}
J.~D.~Scargle, J.~P.~Norris, J.~T.~Bonnell,
 Astrophys.\ J.\  {673} (2008) 972.


\bibitem{Vasileiou}
 V.~Vasileiou {et al.,}
  {Phys.\ Rev.\ D\ } 87 (2013) 122001.

\bibitem{nature090510}
  A.~A.~Abdo {et al.,}
  {Nature} {462} (2009) 331.

\bibitem{PRL090510}
R.~J.~Nemiroff, R.~Connolly, J.~Holmes, A.B.~Kostinski,
  {Phys.\ Rev.\ Lett.\ }  {108} (2012) 231103.





\bibitem{shao}
  L.~Shao, Z.~Xiao, B.-Q.~Ma,
  {Astropart.\ Phys. } {33} (2010) 312.

\bibitem{Amelino}
  G.~Amelino-Camelia, F.~Fiore, D.~Guetta, S.~Puccetti,
 Adv.\ High Energy Phys. 2014 (2014) 597384.

\bibitem{pass8}
  W.~B.~Atwood {et al.,}
  {Astrophys.\ J.\ } {774} (2013) 76.

 \bibitem{e06}
  J.R.~Ellis, N.~E.~Mavromatos, D.~V.~Nanopoulos, A.~S.~Sakharov and E.~K.~G.~Sarkisyan,
  {Astropart.\ Phys.}  { 25} (2006) 402
  [erratum {Astropart.\ Phys.}  { 29}  (2008) 158].

\bibitem[Levan et al.(2013)]{130427Az}
A.~J.~Levan, S.~B.~Cenko, D.~A.~Perley, N.~R.~Tanvir,
{GCN Circ.}
{14455} (2013).

\bibitem{080916Cz}
  J.~Greiner {et al.,}
   {Astron.\ Astrophys.\ } {498} (2009) 89.

\bibitem{090926Az}
  V.~D'Elia {et al.,}
  {Astron.\ Astrophys.\ } {523} (2010) A36.

\bibitem[Cucchiara
\& Fox(2010)]{100414Az}
A.~Cucchiara, D.~B.~Fox,
{GCN Circ.}
{10606} (2010).

\bibitem[Cucchiara et al.(2009)]{090902Bz}
A.~Cucchiara, D.~B.~Fox, N.~Tanvir, E.~Berger,
{GCN Circ.} {9873} (2009).

\bibitem{090510z}
  S.~McBreen {et al.,}
  {Astron.\ Astrophys.\ } {516} (2010) A71.




\bibitem{GBM}
  C.~Meegan {et al.,}
    {Astrophys.\ J.\ } {702} (2009) 791.

\bibitem{Xiao:2009xe}
  Z.~Xiao, B.-Q.~Ma,
  {Phys.\ Rev.\ D } {80} (2009) 116005.

\bibitem{LAT2012}
  M.~Ackermann { et al.}  [Fermi-LAT Collaboration],
  Astrophys.\ J.\ Suppl.\  { 203} (2012) 4.


\bibitem{ApjLswift}
  M.~De Pasquale { et al.}  [Fermi-LAT and GBM Collaborations],
  The Astrophysical Journal Letter { 709} (2010) L146.

\bibitem{1986}
J.~P.~Norris { et al.},
Advances in Space Research 6 (1986) 19.

\bibitem{MKN501}
  J.~Albert { et al.}  [MAGIC Collaboration and other Contributors],
  Phys.\ Lett.\ B { 668} (2008) 253.

\bibitem{darkmatter}
  D.~C.~Latimer,
  Phys.\ Rev.\ D {88} (2013) 063517

\bibitem{Planck-scale}
For discussions, see, e.g., L.~Shao, B.-Q.~Ma,
  Sci.\ China G {\bf 54} (2011) 1771
  [arXiv:1006.3031 [hep-th]], and references therein.



\end{thebibliography}


\end{document}